\documentclass[amsmath,amssymb,jmp,aip,preprint]{revtex4-1}

\usepackage{graphicx}
\usepackage{dcolumn}
\usepackage{bm}

\allowdisplaybreaks
\begin{document}
\sloppy
\newcommand{\rd}{d}
\newcommand{\DDD}[3]{\raisebox{-.6ex}{\scriptsize #1}D_{#2}^{#3}}

\title[Generalized equations of hydrodynamics in fractional derivatives]{Generalized equations of hydrodynamics in fractional derivatives}

\author{P. Kostrobij}
 \affiliation{
Lviv Polytechnic National University, 12 Bandera Str., 79013 Lviv, Ukraine.
}%
\author{B. Markovych}
\email{bohdan.m.markovych@lpnu.ua.}
 \affiliation{
Lviv Polytechnic National University, 12 Bandera Str., 79013 Lviv, Ukraine.
}%
\author{I. Ryzha}
 \affiliation{
Lviv Polytechnic National University, 12 Bandera Str., 79013 Lviv, Ukraine.
}%
\author{M. Tokarchuk}
 \affiliation{%
Institute for Condensed Matter Physics of NAS of Ukraine,\\ 1 Svientsitskii Str., 79011 Lviv, Ukraine.}
 \affiliation{
Lviv Polytechnic National University, 12 Bandera Str., 79013 Lviv, Ukraine.
}%

\begin{abstract}
  We present a general approach for obtaining the generalized transport equations with fractional derivatives using the Liouville equation with fractional derivatives for a system of classical particles and the Zubarev non-equilibrium statistical operator (NSO) method within the Gibbs statistics.
  We obtain the non-Markov equations of hydrodynamics for the non-equilibrium average values of densities of particle number, momentum and energy of liquid in a spatially heterogeneous medium with a fractal structure. For isothermal processes ($\beta=1/k_{B}T =const$), the non-Markov Navier--Stokes equation in fractional derivatives is obtained. We consider models for the frequency dependence of  memory function (viscosity), which lead to the Navier--Stokes equations in fractional derivatives in space and time.
\end{abstract}


\maketitle

\section{Introduction}


 The flows of rheological and non-rheological liquids,
 solutions in porous media are important in almost every field of science and technology~\cite{Mandelbrot1982,Mandelbrot1989,Friedrich1991,Barton199577,Carpinteri1997,Caputo2000693,Falconer2003,Kendel199677,Nonnenmacher1994,Mainardi2010,Fabrizio2014206,Tarasov20161650128,Bouchendouka2022582}. They are relevant in soil science,
 hydrology (ecological problems),
 oil,
 construction industries,
 medicine,
 pharmacology (when making medicines), and others.
 A feature of porous media is their fractal structure.
 Fractals are very important in biological systems. In particular our circulatory system,
 lungs,
 brain,
 arteries,
 veins,
 heartbeat rhythms,
 biological changes of aging and bronchioles are characterized by fractal patterns~\cite{Kendel199677}.

 It is important to note,
 that currently  mathematical direction
 (using fractional calculus~\cite{Oldham2006,Miller1993,Samko1993,Podlubny1998,Mandelbrot1982,Uchaikin2008500,Tarasov2010,Caputo201573}) of modeling sub-,
 super-diffusive,
 electrodiffusive transport processes in porous,
 nanostructured systems with characteristic fractality~\cite{Sahimi1998213,Korosak20071,Hobbie2007,Compte19977277,Metzler20001,Hilfer19951475,Hilfer20003914,Hilfer2000,
Kosztolowicz2005041105,Kosztolowicz2015P10021,Bisquert20002287,Uchaikin2008500,Bisquert2001112,Kosztolowicz2009055004,Pyanylo201484,Kostrobij2016093301,Kostrobij2016163,Glushak201857, Kostrobij201963,Kostrobij201958,Grygorchak2015e,Kostrobij2015154,Grygorchak2017185501,Kostrobij20184099} is quite well developed.
 However,
 it is obvious that for studies of fluid flows in porous media,
 the description of the diffusion level is not enough.
 Therefore,
 from the point of view of theoretical research and mathematical modeling of fluid flows in porous media,
 it is necessary to apply the equations of hydrodynamics,
 which are based on the laws of conservation of mass,
 momentum, and energy densities.
 It is obvious that such equations of hydrodynamics of fluid flows in porous media must take into account their fractal structure.
 Important questions regarding understanding the influence of fractal properties of porous media on fluid flows were considered in many works~\cite{Sahimi2011,Dietrich2005,Winters1994,Almeida19995486,Knudsen2002056310,Huinink2002046301,Balankin2011036310,Lopez2003056314,Stanley200117,Chen2009026301,Tian2006287,Wheatcraft20081377} (and references therein).

 Starting with the works of Tarasov~\cite{Tarasov2005167,Tarasov2005286,Tarasov2010} and~\cite{Ostoja20071085,Li20092521,Balankin2012025302,Balankin2012056314,Li2013057001,Li202020190288} such equations of hydrodynamics of flows in a fractal continuous medium were proposed.
 The peculiarity of these hydrodynamic equations is that they have constant coefficients of viscosity transfer and thermal conductivity,
 and the equations can be used to describe Markov processes.
 In addition,
 in reality,
 memory effects in time and long-range spatial  correlations are important for fluid flows,
 solutions in porous media.
 In this regard,
 it is necessary to obtain hydrodynamic equations taking into account non-Markovianity,
 non-locality and fractality of porous media.
 In this paper,
 for mathematical modeling of fluid flows in porous media,
 we will obtain generalized equations of hydrodynamics with spatio-temporal non-locality.
 In our works~\cite{Kostrobij2016093301,Kostrobij2016163,Glushak201857,Kostrobij201963,Kostrobij201958,Grygorchak2015e,Kostrobij2015154,Grygorchak2017185501,Kostrobij20184099} using the Zubarev non-equilibrium statistical operator (NSO) method~\cite{Zubarev19811509,Zubarev20021,Zubarev20022} and the Liouville equation in fractional derivatives~\cite{Tarasov2010,Tarasov2004123,Tarasov200517,Tarasov2005011102,Tarasov2006033108,Tarasov2006341}, we developed a statistical approach for obtaining generalized transport equations with spatio-temporal non-locality.
 This will be implemented in the Gibbs statistics in the second section.
 In the third section,
 the non-Markov equations of hydrodynamics will be obtained for the non-equilibrium average values of densities of particle number, momentum and energy of fluid in a spatially heterogeneous medium with a fractal structure.
 We will discuss generalized viscosity transfer coefficients,
 thermal conductivity,
 and cross coefficients that describe viscous-thermal correlations.
 In the fourth section, for isothermal processes ($\beta=1/k_{B}T =const$), we will obtain the non-Markov Navier--Stokes equation in fractional derivatives.
 We will consider two models for the frequency dependence of memory (viscosity) function,
 and arrive at Navier--Stokes--type equations in fractional derivatives in space and time for the non-equilibrium average value of particle momentum density.
 It is important to note that the Navier--Stokes equations in fractional derivatives were considered in many works~\cite{Fabrizio2014206,Tarasov20161650128,El-Nabulsi2019449} (and references therein).


 \section{Liouville equation with fractional derivatives for classical system of particles}

 We use the Liouville equation with fractional derivatives obtained by Tarasov in Refs.~\cite{Tarasov2004123,Tarasov200517,Tarasov2005011102,Tarasov2006033108}
 for a non-equilibrium particle function $\rho(x^{N};t)$ of a classical system
 \begin{equation*}
  \frac{\partial }{\partial t}\rho(x^{N};t)+\sum^{N}_{j=1}D^{\alpha}_{\vec{r}_{j}}
  \left(\rho(x^{N};t)\vec{v}_{j}\right) 
  +\sum^{N}_{j=1}D^{\alpha}_{\vec{p}_{j}}
  \left(\rho(x^{N};t)\vec{F}_{j}\right)=0,
 \end{equation*}
 where ${x^{N}=x_{1},\ldots,x_{N}}$,
 ${x_{j}=\{\vec{r}_{j}, \vec{p}_{j}\}}$ are the dimensionless generalized coordinates,
 ${\vec{r}_{j}=(r_{j1},\ldots,r_{jm})}$,
 and generalized momentum,
 ${\vec{p}_{j}=(p_{j1},\ldots,p_{jm})}$,~\cite{Tarasov2006341} of $j$th particle
 in the phase space with a fractional differential volume element~\cite{Kathleen20012203,Tarasov2004123} $d^{\alpha}V=d^{\alpha}x_{1}\ldots d^{\alpha}x_{N}$.
 Here,
 $m=\frac{Mr_{0}}{p_{0}t_{0}}$,
 $M$ is the mass of a particle,
 $r_{0}$ is the characteristic scale in configuration space,
 $p_{0}$  is the characteristic momentum,
 and $t_{0}$ is the characteristic time.
 $d^{\alpha}$ is the fractional differential~\cite{Kathleen20012203} that is defined by
 \[
  d^{\alpha} f(x)=\sum^{2N}_{j=1}D^{\alpha}_{x_{j}}f(x)(dx_{j})^{\alpha}  ,
 \]
 where  $(dx_{j})^{\alpha}:=\mathrm{sgn}(dx_{j})|dx_{j}|^{\alpha}$ ~\cite{Tarasov20082756,Tarasov2011e}
 \begin{equation*}
  D^{\alpha}_{x} f(x)=\frac{1}{\Gamma (n-\alpha)} \int^{x}_{0} \frac{f^{(n)}(z)}{(x-z)^{\alpha +1-n}} dz
 \end{equation*}
 is the Caputo fractional derivative~\cite{Mainardi1997,Caputo1971134,Oldham2006,Samko1993},
 ${n-1<\alpha<n}$,
 $f^{(n)}(z)=\frac{d^{n}}{dz^{n}}f(z)$ with the properties $D^{\alpha}_{x_{j}}1=0$ and $D^{\alpha}_{x_{j}}x_{l}=0$, $(j\neq l)$.
 $\vec{v}_{j}$ are the fields of velocity,
 $\vec{F}_{j}$ is the force field acting on $j$th particle.

 If $\vec{F}_{j}$ does not depend on $\vec{p}_{j}$,
 $\vec{v}_{j}$ does not depend on $\vec{r}_{j}$,  we get
 \[
  \frac{\partial }{\partial t}\rho(x^{N};t)
  +
  \sum^{N}_{j=1}\vec{v}_{j}D^{\alpha}_{\vec{r}_{j}}\rho(x^{N};t)+\sum^{N}_{j=1}\vec{F}_{j}D^{\alpha}_{\vec{p}_{j}}
  \rho(x^{N};t)=0,
 \]
 \[
  \vec{v}_{j}=D^{\alpha}_{\vec{p}_{j}}H(\vec{r},\vec{p}),  \quad
  \vec{F}_{j}=-D^{\alpha}_{\vec{r}_{j}}H(\vec{r},\vec{p}) ,
 \]
 where $H(\vec{r},\vec{p})$ is the Hamiltonian of a system with fractional derivatives~\cite{Tarasov2006341}.
 We get the Liouville equation in the form
 \begin{equation*}
  \frac{\partial }{\partial t}\rho(x^{N};t)+iL_{\alpha}\rho(x^{N};t)=0,
 \end{equation*}
 where $iL_{\alpha}$ is the Liouville operator with the fractional derivatives,
 \begin{equation}\label{eq:2.5}
  iL_{\alpha}\rho(x^{N};t)
  =
  \sum^{N}_{j=1}
  \left[
  D^{\alpha}_{\vec{p}_{j}}H(\vec{r},\vec{p})D^{\alpha}_{\vec{r}_{j}}
  -
  D^{\alpha}_{\vec{r}_{j}}H(\vec{r},\vec{p})D^{\alpha}_{\vec{p}_{j}}\right]
  \rho(x^{N};t).
 \end{equation}

 Solution of the Liouville equation (\ref{eq:2.5}) will be found using the Zubarev non-equilibrium statistical operator method~\cite{Zubarev19811509,Zubarev20021}.
 After choosing parameters of the reduced description,
 taking into account projections we present the non-equilibrium particle function $\rho\left(x^{N};t\right)$
 (as a solution of the Liouville equation) in the general form
 \begin{equation}\label{eq:2.6}
   \rho(x^{N};t)
   =
   \rho_{rel}(x^{N};t)
   -
   \int^{t}_{-\infty}e^{\varepsilon(t'-t)}T(t,t')
   \left(1-P_{rel}(t')\right)iL_{\alpha}\rho_{rel}(x^{N};t')dt',
 \end{equation}
 where
 $T(t,t')=\exp_{+}\left[-\int^{t}_{t'}(1-P_{rel}(t'))iL_{\alpha}dt'\right]$
 is the evolution operator in time containing the projection,
 $\exp_+$ is the ordered exponential,
 ${\varepsilon\to+0}$ after taking the thermodynamic limit,
 $P_{rel}(t')$ is the generalized Kawasaki-Gunton projection operator dependent on a structure of the relevant statistical operator (distribution function),
 $\rho_{rel}(x^{N};t')$.
 By using the Zubarev non-equilibrium statistical operator method~\cite{Zubarev19811509,Zubarev20021,Zubarev20022},
 $\rho_{rel}(x^{N};t')$ will be found from the extremum of the Gibbs entropy
 at fixed values of the observed values $\langle \hat{P}_{n}(x)\rangle^{t}_{\alpha}$,
 taking into account the normalization condition $\langle 1 \rangle^{t}_{\alpha, rel}=1$,
 where the nonequilibrium average values are found respectively~\cite{Tarasov2004123,Tarasov200517,Tarasov2005011102,Tarasov2006033108},
 \begin{equation*}
  \langle \hat{P}_{n} (x)\rangle^{t}_{\alpha}=\hat{I}^{\alpha}(1,\ldots,N)\hat{T}(1,\ldots,N)\hat{P}_{n}\rho(x^{N};t).
 \end{equation*}
 $\hat{I}^{\alpha}(1,\ldots,N)$ has the following form for a system of $N$ particles
 \[
  \hat{I}^{\alpha}(1,\ldots,N)
  =
  \hat{I}^{\alpha}(1),\ldots,\hat{I}^{\alpha}(N),
  \quad
  \hat{I}^{\alpha}(j)=\hat{I}^{\alpha}(\vec{r}_{j})\hat{I}^{\alpha}(\vec{p}_{j})
 \]
 and defines operation of integration
 \begin{equation*}
  \hat{I}^{\alpha}(x)f(x)=\int^{\infty}_{-\infty}f(x)d\mu_{\alpha}(x),
  \quad
  d\mu_{\alpha}(x)=\frac{|x|^{\alpha}}{\Gamma (\alpha)}dx.
 \end{equation*}
 The operator $\hat{T}(1,\ldots,N)=\hat{T}(1),\ldots,\hat{T}(N)$ defines the operation
 \[
  \hat{T}(x_{j})f(x_{j})=\frac{f(\ldots,x'_{j}-x_{j},\ldots)+f(\ldots,x'_{j}+x_{j},\ldots)}{2} .
 \]
 Accordingly,
 the average value,
 which is calculated with the relevant distribution function,
 is defined as
 \[
 \langle (\ldots) \rangle^{t}_{\alpha, rel}=\hat{I}^{\alpha}(1,\ldots,N)\hat{T}(1,\ldots,N)(\ldots)\rho_{rel}(x^{N};t).
 \]
 According to Ref.\cite{Zubarev19811509,Zubarev20021,Zubarev20022},
 from the extremum of the Gibbs entropy  functional
 \begin{align*}
  L_{G}(\rho')&=\ln\hat{I}^{\alpha}(1,\ldots,N)\hat{T}(1,\ldots,N)(\ln(\rho'(t)))\\
  &\quad-\gamma\hat{I}^{\alpha}(1,\ldots,N)\hat{T}(1,\ldots,N)\rho'(t)\\
  &\quad-\sum_{n}\int d\mu_{\alpha}(x)F_{n}(x;t)\hat{I}^{\alpha}(1,\ldots,N)\hat{T}(1,\ldots,N)\hat{P}_{n}(x)\rho'(t)
 \end{align*}
 at fixed values of the observed values $\langle \hat{P}_{n}(x)\rangle^{t}_{\alpha}$
 and
 the condition of normalization $\hat{I}^{\alpha}(1,\ldots,N)\hat{T}(1,\ldots,N)\rho'(t)=1$,
 the relevant distribution function takes the form
 \begin{equation}\label{eq:2.9}
  \rho_{rel}(t)
  =
  \frac{1}{Z_{G}(t)}
  \exp \bigg(
  -\sum_{n}\int d\mu_{\alpha}(x)F_{n}(x;t) \hat{P}_{n}(x)
  \bigg),
  \end{equation}
  $Z_{G}(t)$ is the partition function of the Gibbs distribution,
 which is determined from the normalization condition and has the form
 \begin{equation*}
   Z_{G}(t)=\hat{I}^{\alpha}(1,\ldots,N)\hat{T}(1,\ldots,N)
   \exp\bigg( -
 \sum_{n}\int d\mu_{\alpha}(x)F_{n}(x;t) \hat{P}_{n} (x)\bigg).
 \end{equation*}
 The Lagrangian multiplier $\gamma$ is determined by the normalization condition $\hat{I}^{\alpha}(1,\ldots,N)\hat{T}(1,\ldots,N)\rho'(t)=1$.

 The parameters $F_{n}(x;t)$ are determined from the self-consistency conditions
 \begin{equation*}
   \langle \hat{P}_{n}(x) \rangle^{t}_{\alpha}= \langle \hat{P}_{n}(x) \rangle^{t}_{\alpha, rel}.
 \end{equation*}

 For the general case of parameters $\langle \hat{P}_{n} (x)\rangle_{\alpha }^{t}$ of the reduced description of non-equilibrium processes according to \eqref{eq:2.6} and \eqref{eq:2.9},
 we get the non-equilibrium statistical operator in the form
 \begin{equation} \label{eq:2.12}
  \rho (t)=\rho _{rel} (t)
  +\sum _{n}\int d\mu_{\alpha} (x)
  \int _{-\infty }^{t} e^{\varepsilon (t'-t)} T(t,t')I_{n} (x;t')\rho_{rel} (t') F_{n}(x;t')dt',
 \end{equation}
 where
 \begin{equation*}
   I_{n} (x;t')=\left(1-P(t)\right)iL_{\alpha } \hat{P}_{n} (x)
 \end{equation*}
 are the generalized flows, $P(t)$ is the Mori projection operator, which acts on dynamic variables $A(x)$ and has the following structure:
 \begin{equation*}
 P(t)A(x)=\langle A(x)\rangle^{t}_{\alpha, rel}+\sum _{n}\int d\mu _{\alpha} (x)\frac{\partial \langle A(x)\rangle^{t}_{\alpha, rel}}{\partial \langle \hat{P}_{n} (x)\rangle^{t}}(\hat{P}_{n} (x)- \langle \hat{P}_{n} (x)\rangle^{t}).
 \end{equation*}
 $P(t)$ satisfies the properties:
 $P(t)(1-P(t'))=0$,
 $P(t)\hat{P}_{n}(x)=\hat{P}_{n}(x)$ and is related to the Kawasaki--Gunton operator by the relation:
 \[
  P_{rel}(t)A(x)\rho_{rel}(t)=\rho_{rel}(t)P(t)A(x).
 \]
 By using the non-equilibrium statistical operator~\eqref{eq:2.12},
 we get the generalized transport equation for  parameters $\langle \hat{P}_{n} (x)\rangle_{\alpha}^{t}$ of the reduced description,
 \begin{equation} \label{eq:14}
  \frac{\partial}{\partial t}\langle\hat{P}_{n}(x)\rangle_{\alpha}^{t}
  =\langle iL_{\alpha} \hat{P}_{n} (x)\rangle_{\alpha ,rel}^{t}
  +\sum_{n'}\int d\mu_{\alpha} (x')\int_{-\infty }^{t}e^{\varepsilon (t'-t)}\varphi_{P_{n}P_{n'}}(x,x';t,t') F_{n'} (x';t')dt',
 \end{equation}
 where
 \begin{equation*}
  \varphi_{P_{n} P_{n'}} (x,x';t,t')=\hat{I}^{\alpha} (1,\ldots,N)\hat{T}(1,\ldots,N)
  \left(I_{n} (x;t) T(t,t')I_{n'} (x';t')\rho_{rel} (x^{N} ;t')\right)
 \end{equation*}
 are the generalized transport kernels (the memory functions),
 which describe dissipative processes in the system.
 In the next section,
 we will obtain the generalized equations of hydrodynamics in fractional derivatives.

 \section{Generalized equations of hydrodynamics in fractional derivatives}

 The main observed parameters of the reduced description of the hydrodynamic state of liquids are the average values of densities of particle number  $\langle\hat{n}(\vec{r})\rangle^{t}$,
 momentum $\langle\hat{\vec{p}}(\vec{r})\rangle^{t}$ and energy $\langle\hat{\varepsilon}(\vec{r})\rangle^{t}$,
 where
 \begin{align*}
  \hat{n}(\vec{r})&=\sum_{j=1}^{N}\delta (\vec{r}-\vec{r}_{j}),\\
  \hat{\vec{p}}(\vec{r})&=\sum_{j=1}^{N}\vec{p}_{j}\,\delta (\vec{r}-\vec{r}_{j}),\\
  \hat{\varepsilon}(\vec{r})&=\sum_{j=1}^{N}\frac{p_{j}^{2\alpha}}{2m}\,\delta(\vec{r}-\vec{r}_{j})
  +\frac{1}{2}\sum_{j\neq l=1}^{N}\Phi (\vec{r}_{j},\vec{r}_{l})\,\delta(\vec{r}-\vec{r}_{j})
 \end{align*}
 are the microscopic densities of the number of particles,
 their momentum and total energy.
 $\Phi(\vec{r}_{j},\vec{r}_{l})$ is the pair potential of interaction between particles of a viscous liquid.


 To obtain the generalized transport equations (\ref{eq:14}) for a given set of observed parameters,
 it is necessary to construct the relevant distribution function $\varrho_{rel}(x^N;t)$.
 According to (\ref{eq:2.9}) at fixed values of $\langle\hat{n}(\vec{r})\rangle^{t}$ and
 $\langle \hat{\varepsilon}'(\vec {r})\rangle^{t}$,
 we get the relevant (quasi-equilibrium) distribution in the form:
 \begin{equation}  \label{A1}
  \varrho_{rel}(x^N;t)=\exp\left\{-\Phi (t)-\int  d\mu_{\alpha}(\vec{r})\beta(\vec{r};t)\left(\hat{\varepsilon}'(\vec{r})-\mu(\vec{r};t)\hat{n}(\vec{r})\right)\right\} ,
 \end{equation}
 where
 \begin{equation}  \label{A2}
  \Phi (t)=\ln\hat{I}^{\alpha} (1,\ldots,N)\hat{T}(1,\ldots,N) \exp\left\{-\int  d\mu_{\alpha}(\vec{r})\beta (\vec{r};t)\left(\hat{\varepsilon}'(\vec{r})-\mu(\vec{r};t)\hat{n}(\vec{r})\right)\right\}
 \end{equation}
 is the Massier--Planck functional,
 $\hat{\varepsilon}'(\vec{r})$ is the energy density of particles in the moving system
 \begin{equation*}
  \hat{\varepsilon}'(\vec{r})=\hat{\varepsilon}(\vec{r})-\vec{v}(\vec{r};t)\hat{\vec{p}}(\vec{r})
  +\frac{1}{2}mv^{2}(\vec{r};t)\hat{n}(\vec{r}),
 \end{equation*}
 which moves with a fluid element at an average hydrodynamic velocity
 $\vec{v}(\vec{r};t)=\frac{\langle \hat{\vec{p}}(\vec{r})\rangle ^{t}}{\rho(\vec{r};t)}$,
 where $\rho(\vec{r};t)=m\langle\hat{n} (\vec{r})\rangle^{t}$ is the average mass density.
 At the same time,
 we have:
 \begin{gather*} 
  \hat{\vec{p}}'(\vec{r})=\hat{\vec{p}}(\vec{r})-m \vec{v}(\vec{r};t)\hat{n}(\vec{r}),\\  
  \hat{n}'(\vec{r})=\hat{n}(\vec{r}),\\  
  \vec{r}_{j}=\vec{r}'_{j} ,\quad
  \vec{p}'_{j}=\vec{p}_{j}-m \vec{v}(\vec{r};t).
 \end{gather*}

 The Lagrangian parameters $\beta(\vec{r};t)$, $\mu(\vec{r};t)$ are found from the self-consistency conditions:
 \begin{gather*}
 \langle \hat{\varepsilon}'(\vec{r})\rangle^{t}=\langle \hat{\varepsilon}'(\vec{r})\rangle^{t}_{\alpha,rel},\\
 \langle\hat{n} (\vec{r})\rangle^{t}=\langle\hat{n} (\vec{r})\rangle^{t}_{\alpha, rel}.
 \end{gather*}
 Their physical meaning is determined from the non-equilibrium thermodynamic relationships:
 \begin{equation} \label{A03}
  \frac{\delta \Phi (t)}{\delta \beta(\vec{r};t)}=-\langle \hat{\varepsilon}'(\vec{r})\rangle^{t}, \quad
  \frac{\delta \Phi (t)}{\delta \beta(\vec{r};t)\mu(\vec{r};t)}= \langle\hat{n} (\vec{r})\rangle^{t},
 \end{equation}
 and
 \begin{align}  \label{A4}
  \frac{\delta}{\delta\langle\hat{\varepsilon}'(\vec{r})\rangle^{t}}S(t)&=\beta(\vec{r};t),\\  
  \frac{\delta}{\delta\langle\hat{n}(\vec{r})\rangle^{t}}S(t)&=-\beta(\vec{r};t)\mu(\vec{r};t), \nonumber
 \end{align}
 where
 \begin{equation*}
  S(t)=\Phi(t)
  +\int d\mu_{\alpha}(\vec{r})\beta(\vec{r};t)(\langle\hat{\varepsilon}'(\vec{r})\rangle^{t}
  -\mu(\vec{r};t)\langle\hat{n}(\vec{r})\rangle^{t})
 \end{equation*}
 is the entropy of the non-equilibrium state of the system,
 defined according to Gibbs.
 From the thermodynamic relations (\ref{A03}), (\ref{A4}) it follows that $\beta(\vec{r};t)$ is the inverse value of the non-equilibrium temperature $\beta(\vec{r}; t)=(k_{B}T(\vec{r};t))^{-1}$,
 and $\mu(\vec{r};t)$ is the non-equilibrium value of the chemical potential of  particles.
 Next,
 we define the non-equilibrium thermodynamic potential $\Omega(\vec{r};t)$
 \[
  \Phi(t)=\int d\mu_{\alpha}(\vec{r})\beta(\vec{r};t)\Omega(\vec{r};t),
 \]
 then from (\ref{A2}),
 we find a generalization of the well-known in thermodynamics Gibbs--Duhem relation for the non-equilibrium case:
 \begin{equation}  \label{A06}
  \delta[\beta(\vec{r};t)\Omega(\vec{r};t)]=\langle\hat{\varepsilon}'(\vec{r})\rangle^{t}_{rel}\delta\beta(\vec{r};t)
  -\langle\hat{n}(\vec{r})\rangle^{t}_{rel} \delta[\beta(\vec{r};t)\mu(\vec{r};t)].
 \end{equation}
 We introduce the entropy density $S(\vec{r};t)$ using the equality:
 \[
  \beta^{-1}(\vec{r};t)S(\vec{r};t)=\langle\hat{\varepsilon}'(\vec{r})\rangle^{t}_{\alpha, rel}
  -\langle\hat{n}(\vec{r})\rangle^{t}_{\alpha, rel} \mu(\vec{r};t) - \Omega(\vec{r};t)
 \]
 and rewrite (\ref{A06}) in the form
 \begin{equation*}
  \delta\Omega(\vec{r};t)=S(\vec{r};t)\delta T(\vec{r};t)
  -\langle\hat{n}(\vec{r})\rangle^{t}_{\alpha, rel}\delta\mu(\vec{r};t).
 \end{equation*}
 Then for the variation of the entropy density,
 we obtain  equality:
 \begin{equation*}
  T(\vec{r};t)\delta S(\vec{r};t)=\delta\langle\hat{\varepsilon}'(\vec{r})\rangle^{t}
  -\mu(\vec{r};t)\delta \langle\hat{n}(\vec{r})\rangle^{t}
 \end{equation*}
 is the second law of non-equilibrium thermodynamics in a local form.
 Next,
 we present the relevant distribution (\ref{A1}) in the form:
 \begin{equation} \label{h:010}
  \rho_{rel}(t)=\frac{1}{Z_{rel}(t)}\exp\left(-\int d\mu_{\alpha}(\vec{r})\beta(\vec{r};t)(\hat{\varepsilon}(\vec{r})
  -\hat{\vec{p}}(\vec{r})\vec{v}(\vec{r};t)
  -\nu(\vec{r};t)\hat{n}(\vec{r}))\right),
 \end{equation}
 where
 \begin{multline*} 
  Z_{rel} (t)= \hat{I}^{\alpha} (1,\ldots,N)\hat{T}(1,\ldots,N) \\
  \times\exp\left\{-\int  d\mu_{\alpha}(\vec{r})\beta(\vec{r};t)\left(\hat{\varepsilon} (\vec{r})
  -\hat{\vec{p}}(\vec{r})\vec{v}(\vec{r};t) - \nu(\vec{r};t)\hat{n}(\vec{r})\right)\right\},
 \end{multline*}
 with parameter $\nu(\vec{r};t)=\mu(\vec{r};t)-\frac{1}{2}mv^{2}(\vec{r};t)$.
 Substituting (\ref{h:010}) into (\ref{eq:2.12}),
 as a result,
 for the non-equilibrium distribution function for the hydrodynamic state of the system,
 we obtain:
 \begin{multline}\label{A15}
  \varrho(x^N;t)=\varrho_{rel}(x^N;t) \\
  +\int d\mu_{\alpha}(\vec{r}) \int_{-\infty}^t
  e^{\varepsilon(t'-t)}T_{rel}(t,t')\left(I_{p}(\vec{r};t')\beta(\vec{r};t')\vec{v}(\vec{r};t')
  -I_{\varepsilon}(\vec{r};t')\beta(\vec{r};t')\right)\varrho_{\alpha, rel}(x^N;t')dt',
 \end{multline}
%
 where
 \begin{gather}\label{A16}
  I_{p}(\vec{r};t')=(1-P(t))iL_{\alpha} \hat{\vec{p}}(\vec{r}),\\
  I_{\varepsilon}(\vec{r};t')=(1-P(t))iL_{\alpha} \hat{\varepsilon}(\vec{r})\nonumber
 \end{gather}
 are the generalized flows describing viscous and thermal processes,
 and $I_{n}(\vec{r};t')=(1-P(t))iL_{\alpha} \hat{n}(\vec{r})=0$,
 $P(t)$ is the generalized Mori projection operator that has the following structure:
 \begin{multline*}
  P(t)A=\langle A\rangle_{\alpha, rel}^t+\int d\mu_{\alpha}(\vec{r})\frac{\delta
  \langle A\rangle _{\alpha, rel}^t}{\delta \langle \hat{n}(\vec{r})\rangle^t}(\hat{n}(\vec{r})-\langle
  \hat{n}(\vec{r})\rangle^t) \\
  + \int d\mu_{\alpha}(\vec{r})\frac{\delta
  \langle A\rangle _{\alpha, rel}^t}{\delta \langle \hat{\vec{p}}(\vec{r})\rangle^t}(\hat{\vec{p}}(\vec{r})-\langle
  \hat{\vec{p}}(\vec{r})\rangle^t)
  +\int d\mu_{\alpha}(\vec{r})\frac{\delta\langle A\rangle _{\alpha, rel}^t}{\delta \langle \hat{\varepsilon}(\vec{r})\rangle^t}(\hat{\varepsilon}(\vec{r})-\langle
  \hat{\varepsilon}(\vec{r})\rangle^t) .
 \end{multline*}
 $P(t)$ has the following properties:
 $P(t)\hat{p}_{n}(\vec{r})=\hat{p}_{n}(\vec{r})$,
 (where $\hat{p}_{n}(\vec{r})=\hat{n}(\vec{r}), \hat{\vec{p}}(\vec{r}), \hat{\varepsilon}(\vec{r})$),
 $P(t)(1-P(t'))=0$.
 The non-equilibrium distribution function (\ref{A15}) has two components:
 $\rho_{rel}$ describes the non-dissipative processes,
 and the next two terms describe the non-Markov non-equilibrium processes of momentum and energy transfer in a system with a fractal structure.
 The non-equilibrium distribution function (\ref{A15}) makes it possible to obtain the transfer equation for
 $\langle\hat{n}(\vec{r})\rangle^{t}$,
 $\langle\hat{\vec{p}}(\vec{r})\rangle^{t}$ and $\langle\hat{\varepsilon}(\vec{r})\rangle^{t}$,
 which are the generalized equations of hydrodynamics.
 If we consider the following equations
 \[
  \frac{\partial }{\partial t}\langle\hat{p}_{n}(\vec{r})\rangle^{t}= \langle iL_{\alpha}\hat{p}_{n}(\vec{r})\rangle^{t}
 \]
 and identities
 \[
  \langle (1-P(t))iL_{\alpha}\hat{p}_{n}(\vec{r})\rangle^{t}
  =
  \langle iL_{\alpha}\hat{p}_{n}(\vec{r})\rangle^{t}-\langle iL_{\alpha}\hat{p}_{n}(\vec{r})\rangle_{\alpha, rel}^{t},
 \]
 using (\ref{A15}),
 for $\langle \hat{n}(\vec{r})\rangle^{t}$,
 $\langle \hat{\vec{p}}(\vec{r})\rangle^{t}$ and $\langle \hat{\varepsilon}(\vec{r})\rangle^{t}$ we obtain the generalized equations of hydrodynamics
 \begin{align*} 
  \frac{\partial}{\partial t}\left\langle\hat{n}(\vec{r})\right\rangle^{t}
  &=-\frac{1}{m}D^{\alpha}_{\vec{r}}\cdot\langle\hat{\vec{p}}(\vec{r})\rangle^{t},\\ 
  \frac{\partial}{\partial t}\langle\hat{\vec{p}}(\vec{r})\rangle^{t}
  &=\langle iL_{\alpha}\hat{\vec{p}}(\vec{r})\rangle_{\alpha,rel}^{t}
    -\int d\mu_{\alpha}(\vec{r}')\int_{-\infty}^{t}e^{\varepsilon(t'-t)}
    \phi_{pp}(\vec{r},\vec{r}';t,t')\,\beta(\vec{r}';t')\vec{v}(\vec{r}';t')dt'\\
  &\quad+\int d\mu_{\alpha}(\vec{r}')\int_{-\infty}^{t}e^{\varepsilon(t'-t)}
    \phi_{p \varepsilon}(\vec{r},\vec{r}';t,t')\,\beta(\vec{r}';t')dt',\nonumber\\ 
  \frac{\partial}{\partial t}\langle\hat{\varepsilon}(\vec{r})\rangle^{t}
  &=\langle iL_{\alpha}\hat{\varepsilon} (\vec{r})\rangle_{\alpha,rel}^{t}
    -\int d\mu_{\alpha}(\vec{r}')\int_{-\infty}^{t}e^{\varepsilon(t'-t)}
  \phi_{\varepsilon p}(\vec{r},\vec{r}';t,t')\,\beta(\vec{r}';t')\vec{v}(\vec{r}';t')dt'\\
  &\quad+\int d\mu_{\alpha}(\vec{r}')\int_{-\infty}^{t}e^{\varepsilon(t'-t)}
  \phi_{\varepsilon \varepsilon}(\vec{r},\vec{r}';t,t')\,\beta(\vec{r}';t')dt',\nonumber
 \end{align*}
 where
 \begin{align*} 
  \phi_{pp}(\vec{r},\vec{r}';t,t')\,\beta(\vec{r}';t')&=\hat{I}^{\alpha} (1,\ldots,N)\hat{T}(1,\ldots,N)
  (I_{p} (\vec{r};t) T(t,t')I_{p} (\vec{r}';t')\rho_{rel} (x^{N} ;t')),\\ 
  \phi_{p\varepsilon}(\vec{r},\vec{r}';t,t')\,\beta(\vec{r}';t')&=\hat{I}^{\alpha} (1,\ldots,N)\hat{T}(1,\ldots,N)
  (I_{p} (\vec{r};t) T(t,t')I_{\varepsilon} (\vec{r}';t')\rho_{rel} (x^{N} ;t')),\\ 
  \phi_{\varepsilon p}(\vec{r},\vec{r}';t,t')\,\beta(\vec{r}';t')&=\hat{I}^{\alpha} (1,\ldots,N)\hat{T}(1,\ldots,N)
  (I_{\varepsilon} (\vec{r};t) T(t,t')I_{p} (\vec{r}';t')\rho_{rel} (x^{N} ;t')),\\ 
  \phi_{\varepsilon \varepsilon}(\vec{r},\vec{r}';t,t')\,\beta(\vec{r}';t')&=\hat{I}^{\alpha} (1,\ldots,N)\hat{T}(1,\ldots,N)
  (I_{\varepsilon} (\vec{r};t) T(t,t')I_{\varepsilon} (\vec{r}';t')\rho_{rel} (x^{N} ;t'))
 \end{align*}
 are the generalized memory functions
 (generalized transfer kernels)
 that are associated with the corresponding generalized coefficients of viscosity,
 thermal conductivity,
 and cross coefficients describing viscous-thermal transfer processes for a system with a fractal structure.
 To establish such a relation,
 it is necessary to reveal the action of the Liouville operator $iL_{\alpha}$ on $\hat{\vec{p}}(\vec{r})$ and $\hat{\varepsilon} (\vec{r})$ in generalized flows~(\ref{A16}).
 The result of the Liouville operator is as follows:
 \begin{align*} 
  iL_{\alpha}\hat{\vec{p}}(\vec{r})
  &=\sum^{N}_{j=1}D^{\alpha}_{\vec{p}_{j}}H(\vec{r},\vec{p})D^{\alpha}_{\vec{r}_{j}}\hat{\vec{p}}(\vec{r})
  -\sum^{N}_{j=1}D^{\alpha}_{\vec{r}_{j}}H(\vec{r},\vec{p})D^{\alpha}_{\vec{p}_{j}}\hat{\vec{p}}(\vec{r})
  =-D^{\alpha}_{\vec{r}}:\overleftrightarrow{T}(\vec{r}), \\ 
  iL_{\alpha}\hat{\varepsilon}(\vec{r})
  &=\sum^{N}_{j=1}D^{\alpha}_{\vec{p}_{j}}H(\vec{r},\vec{p})D^{\alpha}_{\vec{r}_{j}}\hat{\varepsilon}(\vec{r})
  -\sum^{N}_{j=1}D^{\alpha}_{\vec{r}_{j}}H(\vec{r},\vec{p})D^{\alpha}_{\vec{p}_{j}}\hat{\varepsilon}(\vec{r})
  =-D^{\alpha}_{\vec{r}}\cdot\hat{\vec{j}}_{\varepsilon}(\vec{r}),
 \end{align*}
 where the structure of the stress tensor density $\overleftrightarrow{T}(\vec{r})$ and the energy flow density $\hat{\vec{j}}_{\varepsilon}(\vec{r})$ depends on the form of the Hamiltonian $H(\vec{r},\vec{p})$ for a system with a fractal structure.
 Then the generalized fluxes (\ref{A16}) will be expressed in terms of the generalized viscous stress tensor $\overleftrightarrow{T}(\vec{r})$ and the generalized energy flux $\hat{\vec{j}}_{\varepsilon}(\vec{r})$,
 respectively,
 and will have the form:
 \begin{align}\label{h:10}
  I_{p}(\vec{r};t')&=-(1-P(t))D^{\alpha}_{\vec{r}}:\overleftrightarrow{T}(\vec{r}),\\
  I_{\varepsilon}(\vec{r};t')&=-(1-P(t))D^{\alpha}_{\vec{r}}\cdot  \hat{\vec{j}}_{\varepsilon}(\vec{r}).\nonumber
 \end{align}
 Then,
 according to the structure of generalized flows (\ref{h:10}),
 the generalized equations of hydrodynamics in fractional derivatives have the form:
 \begin{align}\label{h:011}
  \frac{\partial}{\partial t}\left\langle\hat{n}(\vec{r})\right\rangle^{t}
  &=-\frac{1}{m} D^{\alpha}_{\vec{r}} \cdot  \langle \hat{\vec{p}} (\vec{r})\rangle^{t},\\ \label{h:022}
  \frac{\partial}{\partial t}\langle\hat{\vec{p}}(\vec{r})\rangle^{t}
  &=-D^{\alpha}_{\vec{r}}: \langle \overleftrightarrow{T}(\vec{r}) \rangle_{\alpha,rel}^{t} \\
  &\quad+
  \int d\mu_{\alpha}(\vec{r}')
  \int_{-\infty}^{t}e^{\varepsilon(t'-t)}D^{\alpha}_{\vec{r}}:
  \overleftrightarrow{\eta}(\vec{r},\vec{r}';t,t'):D^{\alpha}_{\vec{r}'}\,\beta(\vec{r}';t')\vec{v}(\vec{r}';t')dt' \nonumber \\
  &\quad-\int d\mu_{\alpha}(\vec{r}')
  \int_{-\infty}^{t}e^{\varepsilon(t'-t)}
  D^{\alpha}_{\vec{r}}:
  \xi_{\pi q}(\vec{r},\vec{r}';t,t')\cdot D^{\alpha}_{\vec{r}'}\,\beta(\vec{r}';t')dt',\nonumber \\ \label{h:033}
  \frac{\partial}{\partial t}\langle\hat{\varepsilon}(\vec{r})\rangle^{t}
  &=-D^{\alpha}_{\vec{r}}\cdot\langle \hat{\vec{j}}_{\varepsilon}(\vec{r})\rangle_{\alpha,rel}^{t} \\
  &\quad+\int d\mu_{\alpha}(\vec{r}')\int_{-\infty}^{t}e^{\varepsilon(t'-t)}D^{\alpha}_{\vec{r}}\cdot
  \xi_{q \pi}(\vec{r},\vec{r}';t,t'): D^{\alpha}_{\vec{r}'}\,\beta(\vec{r}';t')\vec{v}(\vec{r}';t')dt' \nonumber \\
  &\quad-\int d\mu_{\alpha}(\vec{r}')\int_{-\infty}^{t}e^{\varepsilon(t'-t)}D^{\alpha}_{\vec{r}}\cdot \lambda(\vec{r},\vec{r}';t,t')\cdot D^{\alpha}_{\vec{r}'}\,\beta(\vec{r}';t')dt',\nonumber
 \end{align}
 where
 \begin{equation} \label{h:034}
  \overleftrightarrow{\eta}(\vec{r},\vec{r}';t,t')
  =\left\langle\overleftrightarrow{\pi}(\vec{r};t) T_{rel}(t,t')\overleftrightarrow{\pi}(\vec{r}';t')\right\rangle_{\alpha,rel}^{t'}
 \end{equation}
 is the tensor of the generalized viscosity coefficient,
 \begin{align} \label{h:035}
  \xi_{\pi q}(\vec{r},\vec{r}';t,t')
  &=\langle \overleftrightarrow{\pi}(\vec{r};t)T_{rel}(t,t')\vec{q}(\vec{r}';t')\rangle_{\alpha,rel}^{t'},\\ \label{h:036}
  \xi_{q \pi}(\vec{r},\vec{r}';t,t')
  &=\langle \vec{q}(\vec{r};t)T_{rel}(t,t')\overleftrightarrow{\pi}(\vec{r}';t')\rangle_{\alpha,rel}^{t'}
 \end{align}
 are the cross-generalized transport coefficients describing dissipative viscous-thermal dynamic correlations,
 \begin{equation} \label{h:037}
  \lambda(\vec{r},\vec{r}';t,t')=\langle\vec{q}(\vec{r};t)T_{rel}(t,t')\vec{q}(\vec{r}';t')\rangle_{\alpha,rel}^{t'}
 \end{equation}
 is the generalized thermal conductivity coefficient of particles of a system with a fractal structure.
 These generalized transport coefficients are time correlation functions built on the generalized densities of the viscous stress tensor $\overleftrightarrow{\pi} (\vec{r};t)$ and the energy flow $\vec{q}(\vec{r};t)$
 \begin{align} \label{h:038}
  \overleftrightarrow{\pi}(\vec{r};t)&=(1-P(t))\overleftrightarrow{T}(\vec{r}),\\ \label{h:039}
  \vec{q}(\vec{r};t)&=(1-P(t))\hat{\vec{j}}_{\varepsilon}(\vec{r}),
 \end{align}
 with the averaging operation on the relevant distribution $\rho_{rel}(t)$ (\ref{h:010}).
 $\overleftrightarrow{\pi}(\vec{r};t)$ is the difference of the density of the viscous stresses tensor  $\overleftrightarrow{T}(\vec{r})$ and the projection $P(t)\overleftrightarrow{T }(\vec{r})$ of densities of particle number  $\langle\hat{n} (\vec{r})\rangle^{t}$,
 momentum $\langle \hat {\vec{p}}(\vec{r})\rangle^{t}$ and energy $\langle \hat{\varepsilon}(\vec{r})\rangle^{t}$ onto the space of the main parameters of the reduced description of the hydrodynamic state of fluid.
 Similarly,
 $\vec{q}(\vec{r};t)$ is the difference of the energy flow density $\hat{\vec{j}}_{\varepsilon}(\vec{r})$ and the projection $P( t)\hat{\vec{j}}_{\varepsilon}(\vec{r})$ onto the same space of the main parameters of the reduced description of the hydrodynamic state of fluid.

 Therefore,
 according to the structure $\overleftrightarrow{\pi}(\vec{r};t)$ and $\vec{q}(\vec{r};t)$,
 these dynamic variables change faster in time than dynamic variables $\hat{n}(\vec{r})$,
 $\hat {\vec{p}}(\vec{r})$,
 $\hat{\varepsilon}(\vec{r})$.
 The obtained system of generalized equations of hydrodynamics~(\ref{h:011}), (\ref{h:022}), (\ref{h:033}) takes into account memory effects in time (non-Markov processes) and non-locality in space.
 The generalized transfer coefficients~(\ref{h:034})--(\ref{h:037}) contain mechanisms of transfer phenomena in a system with a fractal structure.
 Calculation of these generalized transfer coefficients (time correlation functions of dynamic variables $\overleftrightarrow{\pi}(\vec{r};t)$ and $\vec{q}(\vec{r};t)$) is a separate important task.
 In particular,
 in the approximation of Markov processes in time and neglecting non-locality in spatial coordinates,
 we obtain:
 \begin{equation} \label{h:040}
  \overleftrightarrow{\eta}(\vec{r},\vec{r}';t,t')=\eta\delta(t-t')\delta(\vec{r}-\vec{r}'), \quad
  \lambda(\vec{r},\vec{r}';t,t')=\lambda\delta (t-t')\delta(\vec{r}-\vec{r}'),
 \end{equation}
 while the cross transfer coefficients $\xi_{\pi q}(\vec{r},\vec{r}';t,t')=0$,
 $\xi_{q \pi}(\vec{r},\vec{r}';t,t')=0$,
 where $\eta$ and $\lambda$ are the viscosity and thermal conductivity transfer coefficients for the system as a continuous medium with a fractal structure.
 In the Markov approximation~(\ref{h:040}) for transfer coefficients,
 the hydrodynamic equations~(\ref{h:034})--(\ref{h:037}) are equivalent to the equations obtained by Tarasov~\cite{Tarasov2005167,Tarasov2005286,Tarasov2010} and others~\cite{Ostoja20071085,Li20092521,Balankin2012025302,Balankin2012056314,Li2013057001}.
 It is important to note that the contribution of cross transfer coefficients $\xi_{\pi q}(\vec{r},\vec{r}';t,t')$,
 $\xi_{q \pi}(\vec{r},\vec{r}';t,t')$ is important for intermediate values of dependence on coordinates and time
 (or, in the language of wave vectors $\vec{k}$ and frequencies $\omega$).
 This was shown for various systems,
 in particular,
 for liquids~\cite{Akcasu1970962,Schepper1988271,Mryglod1995235},
 for semiquantum helium~\cite{Ignatyuk1999295,Ignatyuk1999857},
 ionic~\cite{Zubarev1987164} systems,
 and others.

 \section{Navier--Stokes equation in fractional derivatives with memory}

 In the case $\beta(\vec{r};t)=\beta$ (isothermal processes),
 the equations of generalized hydrodynamics will have the following form:
 \begin{align} 
  \frac{\partial}{\partial t}\left\langle\hat{n}(\vec{r})\right\rangle^{t}
  &=-\frac{1}{m} D^{\alpha}_{\vec{r}} \cdot  \langle \hat{\vec{p}} (\vec{r})\rangle^{t},\nonumber\\ \label{h:0222}
  \frac{\partial}{\partial t}\langle \hat{\vec{p}} (\vec{r})\rangle^{t}
  &=-D^{\alpha}_{\vec{r}}: \langle \overleftrightarrow{T}(\vec{r}) \rangle_{\alpha,rel}^{t} \\
  &\quad+\int d\mu_{\alpha}(\vec{r}')\int_{-\infty}^{t}e^{\varepsilon(t'-t)}D^{\alpha}_{\vec{r}}:
  \overleftrightarrow{\eta}(\vec{r},\vec{r}';t,t'):D^{\alpha}_{\vec{r}'}\,\beta(\vec{r}';t')\vec{v}(\vec{r}';t')dt'.\nonumber
 \end{align}
 This system of equations describes viscoelastic processes and takes into account the spatial fractality of the system and memory effects in the generalized particle viscosity coefficient $\overleftrightarrow{\eta}(\vec{r},\vec{r}';t,t' )$ in the Gibbs statistics.
 The spatial fractality of the system obviously affects the processes of particle transport,
 which can appear as temporal multifractality with characteristic relaxation times.
 It is known that non-equilibrium correlation functions $\overleftrightarrow{\eta}(\vec{r},\vec{r}';t,t')$ cannot be calculated exactly,
 so certain approximations are used based on physical considerations.
 In the time interval $-\infty \div t$,
 particle transport processes in a spatially heterogeneous system can be characterized by a set of relaxation times,
 which are related to the nature of particles interaction with particles of a medium with a fractal structure.
 To reveal temporal multifractality in the generalized Navier--Stokes equation~(\ref{h:0222}),
 we use the following approximation for the generalized particle viscosity coefficient:
 \begin{equation*}
  \overleftrightarrow{\eta}(\vec{r},\vec{r}';t,t')=W_{\eta}(t,t')\overleftrightarrow{\eta}(\vec{r},\vec{r}'),
 \end{equation*}
 where $W_{\eta}(t,t')$ can be defined as the function of memory in time.
 Taking this into account,
 the equation~(\ref{h:0222}) can be presented in the form:
 \begin{equation}\label{eq:2.191}
  \frac{\partial}{\partial t}\langle\hat{\vec{p}}(\vec{r})\rangle^{t}
  +D^{\alpha}_{\vec{r}}:\langle\overleftrightarrow{T}(\vec{r})\rangle_{\alpha,rel}^{t}
  =\int_{-\infty}^{t}e^{\varepsilon(t'-t)}W_{\eta}(t,t')(t,t')\bar{\eta}(\vec{r};t')dt',
 \end{equation}
 where
 \begin{equation*}
  \bar{\eta}(\vec{r};t')=\int d\mu_{\alpha}(\vec{r}')D^{\alpha}_{\vec{r}}:
  \overleftrightarrow{\eta}(\vec{r},\vec{r}):D^{\alpha}_{\vec{r}'}\,\beta\vec{v}(\vec{r}';t').
 \end{equation*}

 Next,
 we apply the Fourier transform to  equation~(\ref{eq:2.191}),
 as a result,
 in the frequency image we get:
 \begin{equation}\label{eq:2.193}
  i\omega\langle\hat{\vec{p}}(\vec{r})\rangle^{\omega}
  +D^{\alpha}_{\vec{r}}:
  \langle\overleftrightarrow{T}(\vec{r})\rangle_{\alpha,rel}^{\omega}
  =W_{\eta}(\omega)\bar{\eta}(\vec{r};\omega).
 \end{equation}
 We present the frequency dependence of the memory function in the form:
 \begin{equation*}
  W_{\eta}(\omega)=\frac{1}{(i\omega)^{\xi-1}}, \quad
  0<\xi\leqslant1.
 \end{equation*}
 Then  equation (\ref{eq:2.193}) can be presented in the form:
 \begin{equation}\label{eq:2.1095}
  (i\omega)^{1-\xi}i\omega\langle\hat{\vec{p}}(\vec{r})\rangle^{\omega}+(i\omega)^{1-\xi}D^{\alpha}_{\vec{r}}:
  \langle\overleftrightarrow{T}(\vec{r})\rangle_{\alpha,rel}^{\omega}
  =(i\omega)^{1-\xi}\bar{\eta}(\vec{r};\omega).
 \end{equation}
 Next,
 we will use the Fourier transform to fractional derivatives of functions:
 \begin{equation*}
  L\left(\DDD{0}{t}{1-\xi}f(t);i\omega\right)=(i\omega)^{1-\xi} L(f(t);i\omega),
 \end{equation*}
 where
 \[
  \DDD{0}{t}{1-\xi}f(t)=\frac{1}{\Gamma(\xi)}\frac{\rd}{\rd t}\int^{t}_{0}
  \frac{f(\tau)}{(t-\tau)^{1-\xi}}\,\rd\tau
 \]
 is the Riemann--Liouville fractional derivative.
 Using it,
 the inverse transform in equation~(\ref{eq:2.1095}) to the time dependence gives the generalized equation of the Navier--Stokes type for describing viscoelastic processes taking into account spatial fractality:
 \begin{equation}\label{eq:2.1906}
  \DDD{0}{t}{\xi}\big\langle\hat{\vec{p}}(\vec{r})\big\rangle^{t}
  +
  \DDD{0}{t}{\xi-1} D^{\alpha}_{\vec{r}}: \big\langle \overleftrightarrow{T}(\vec{r})\big\rangle_{\alpha,rel}^{t}
  =\int d\mu_{\alpha}(\vec{r}')D^{\alpha}_{\vec{r}}:
  \overleftrightarrow{\eta}(\vec{r},\vec{r}):D^{\alpha}_{\vec{r}'}\,\beta\vec{v}(\vec{r}';t).
 \end{equation}
 The equation \eqref{eq:2.1906} contains significant spatial heterogeneity in $\overleftrightarrow{\eta}(\vec r,\vec r')$.
 If spatial heterogeneity is neglected:
 \begin{equation}\label{eq:2.197}
  \overleftrightarrow{\eta}(\vec r,\vec r')=\eta\delta(\vec r-\vec r'),
 \end{equation}
 then we will get
 \begin{equation*}
  \DDD{0}{t}{\xi}\langle\hat{\vec{p}}(\vec{r})\rangle^{t}
  +
  \DDD{0}{t}{\xi-1}D^{\alpha}_{\vec{r}}: \langle \overleftrightarrow{T}(\vec{r}) \rangle_{\alpha,rel}^{t}
  =
  \eta \,D^{\alpha}_{\vec{r}}:\overleftrightarrow{I}:D^{\alpha}_{\vec{r}'}\,\beta\vec{v}(\vec{r};t),
 \end{equation*}
 where $\overleftrightarrow{I}$ is the unit tensor.
 Next,
 we present the frequency dependence of the memory function in the form,
 with  introduction of the relaxation time $\tau$
 (which characterizes the processes of particle transport in the system):
 \begin{equation*}
  W_{\eta}(\omega)=\frac{(i\omega)^{1-\xi}}{1+i\omega \tau}, \quad
  0<\xi\leqslant1.
 \end{equation*}
 Then  equation (\ref{eq:2.193}) can be presented in the form:
 \begin{equation}\label{eq:2.195}
  (1+i\omega\tau)i\omega\big\langle\hat{\vec{p}}(\vec{r})\big\rangle^{\omega}
  +(1+i\omega\tau)D^{\alpha}_{\vec{r}}:\big\langle \overleftrightarrow{T}(\vec{r})\big\rangle_{\alpha,rel}^{\omega} =(i\omega)^{1-\xi}\bar{\eta}(\vec{r};\omega).
 \end{equation}

 Using it,
 the inverse transform in  equation (\ref{eq:2.195}) to the time dependence gives the generalized equation of the Cattaneo--Navier--Stokes type for describing viscoelastic processes taking into account spatial fractality:
 \begin{equation*}
  \tau\frac{\partial^{2}}{\partial t^{2}}\langle\hat{\vec{p}}(\vec{r})\rangle^{t}
  +\frac{\partial}{\partial t}\langle \hat{\vec{p}}(\vec{r})\rangle^{t}
  +\left(1+\tau\frac{\partial}{\partial t}\right)D^{\alpha}_{\vec{r}}:
   \langle\overleftrightarrow{T}(\vec{r})\rangle_{\alpha,rel}^{t}
   =\DDD{0}{t}{1-\xi}\bar{\eta}(\vec{r};t)
 \end{equation*}
 or in expanded form:
 \begin{multline*}
  \tau\frac{\partial^{2}}{\partial t^{2}}\big\langle\hat{\vec{p}}(\vec{r})\big\rangle^{t}
  +\frac{\partial}{\partial t}\langle\hat{\vec{p}}(\vec{r})\rangle^{t}
  +\left(1+\tau\frac{\partial}{\partial t}\right)
  D^{\alpha}_{\vec{r}}:\big\langle\overleftrightarrow{T}(\vec{r})\big\rangle_{\alpha,rel}^{t} \\
  =\DDD{0}{t}{1-\xi}\int d\mu_{\alpha}(\vec{r}')D^{\alpha}_{\vec{r}}:
  \overleftrightarrow{\eta}(\vec{r},\vec{r}):D^{\alpha}_{\vec{r}'}\,\beta\vec{v}(\vec{r}';t)
 \end{multline*}
 is the generalized equation of the Navier--Stokes type for describing viscoelastic processes in the Gibbs statistics with multitemporal and spatial fractality.
 If spatial heterogeneity is neglected~\eqref{eq:2.197},
 we will get
 \begin{equation*}
  \tau\frac{\partial^{2}}{\partial t^{2}}\big\langle\hat{\vec{p}}(\vec{r})\big\rangle^{t}
  +
  \frac{\partial}{\partial t}\big\langle\hat{\vec{p}}(\vec{r})\big\rangle^{t}
  +
  \left(1+\tau\frac{\partial}{\partial t}\right)D^{\alpha}_{\vec{r}}:
  \big\langle\overleftrightarrow{T}(\vec{r})\big\rangle_{\alpha,rel}^{t}
  =\DDD{0}{t}{1-\xi}\eta\,D^{\alpha}_{\vec{r}}:
  \overleftrightarrow{I}:D^{\alpha}_{\vec{r}'}\,\beta\vec{v}(\vec{r};t)
 \end{equation*}
 is the Navier--Stokes equation with spatial and temporal fractality with constant viscosity coefficients in the Gibbs statistics.

\section{Conclusions}

 We present the general approach for obtaining the generalized transport equations with the fractional derivatives by using the Liouville equation with the fractional derivatives~\cite{Tarasov2004123,Tarasov200517,Tarasov2005011102,Tarasov2006033108} for a system of classical particles and the Zubarev non-equilibrium statistical operator method within the Gibbs statistics.
 In this approach, we obtain
 the  non-Markov equations of hydrodynamic in a spatially non-homogeneous medium with a fractal structure.
 We analyze the structure of generalized viscosity and thermal conductivity transfer coefficients,
  and cross transfer coefficients.
 In the Markov approximation (there is no memory in time) and neglecting spatial heterogeneity,
 the generalized equations of hydrodynamics can be reduced to Tarasovs equations of hydrodynamics~\cite{Tarasov2005167,Tarasov2005286,Tarasov2010} for a continuous medium with a fractal structure.


 In the case of isothermal processes,
 we obtain a system of equations describing the non-Markov viscoelastic processes.
 It takes into account spatial fractality and memory effects in the generalized fluid viscosity coefficient.
 It is obvious that the spatial fractality of the system affects the processes of particle transport,
 and this influence,
 in particular,
 can appear as temporal multifractality with characteristic relaxation times.
 In this regard,
 we model the non-Markovianity of viscous processes in time by choosing the frequency dependence of the memory function in the form $\frac{1}{(i\omega)^{\xi-1}}$,
 $0<\xi\leqslant1$ and $\frac{(i\omega)^{1-\xi}}{1+i\omega \tau}$,
 $0<\xi\leqslant1$ with characteristic relaxation time $\tau$.
 Using the fractional calculus method for these non-Markovian models,
 we obtain the Navier--Stokes--type equations in fractional derivatives with nonlocality of space-time.
 The first and important further step is to study the frequency spectrum of the given equations,
 which contain $D^{\alpha}_{\vec{r}}:\langle \overleftrightarrow{T}(\vec{r})\rangle_{\alpha,rel}^{t}$.
 It is essential to note that in $\langle \overleftrightarrow{T}(\vec{r})\rangle_{\alpha,rel}^{t}$ the averaging operation is performed with the known relevant statistical operator $\rho_{rel}(t)$~\eqref{A1}.
 The statistical sum of the relevant operator can basically be calculated at the microscopic level,
 taking into account the nature of system particles interaction.
 The method of calculation can be the method of collective variables~\cite{Tokarchuk2022440}.
 In the following works,
 we will consider these questions and problems of solutions of the Navier--Stokes type equation in fractional derivatives.


\begin{thebibliography}{10}
\expandafter\ifx\csname url\endcsname\relax
  \def\url#1{\texttt{#1}}\fi
\expandafter\ifx\csname urlprefix\endcsname\relax\def\urlprefix{URL }\fi
\expandafter\ifx\csname href\endcsname\relax
  \def\href#1#2{#2} \def\path#1{#1}\fi

\bibitem{Mandelbrot1982}
B.~B. Mandelbrot, The fractal geometry of nature, W. H. Freeman and Company,
  1982.

\bibitem{Mandelbrot1989}
C.~H. Scholz, B.~B. Mandelbrot, Fractals in geophysics, Basel: Birkh\"{a}user
  Verlag, 1989.

\bibitem{Friedrich1991}
C.~Friedrich, Relaxation functions of rheological constitutive equations with
  fractional derivatives: Thermodynamical constraints, in: J.~Casas-V\'azquez,
  D.~Jou (Eds.), Rheological Modelling: Thermodynamical and Statistical
  Approaches, Springer Berlin Heidelberg, Berlin, Heidelberg, 1991, pp.
  321--330.

\bibitem{Barton199577}
C.~C. Barton, P.~R. Pointe (Eds.), Fractals in the Earth Sciences, Springer US,
  Boston, MA, 1995.
\newblock \href {http://dx.doi.org/https://doi.org/10.1007/978-1-4899-1397-5}
  {\path{doi:https://doi.org/10.1007/978-1-4899-1397-5}}.

\bibitem{Carpinteri1997}
A.~Carpinteri, F.~Mainardi (Eds.), {Fractals and Fractional Calculus in
  Continuum Mechanics}, Springer Vienna, Vienna, 1997.
\newblock \href {http://dx.doi.org/https://doi.org/10.1007/978-3-7091-2664-6}
  {\path{doi:https://doi.org/10.1007/978-3-7091-2664-6}}.

\bibitem{Caputo2000693}
M.~Caputo, Models of flux in porous media with memory, Water Resources Research
  36~(3) (2000) 693--705.
\newblock \href {http://dx.doi.org/https://doi.org/10.1029/1999WR900299}
  {\path{doi:https://doi.org/10.1029/1999WR900299}}.

\bibitem{Falconer2003}
K.~J. Falconer, Fractal geometry-mathematical foundations and applications,
  Wiley, New York, NY, 2003.

\bibitem{Kendel199677}
N.~C. Kendel, D.~J. Walker, Fractals in the biological sciences, COENOSES 11
  (1996) 77--100.

\bibitem{Nonnenmacher1994}
T.~F. Nonnenmacher, G.~A. Losa, E.~R. Weibel (Eds.), Fractals in Biology and
  Medicine (Mathematics and Biosciences in Interaction), Birkh\"{a}user, 1994.

\bibitem{Mainardi2010}
F.~Mainardi, Fractional Calculus and Waves in Linear Viscoelasticity, Imperial
  College Press, 2010.
\newblock \href {http://dx.doi.org/https://doi.org/10.1142/p614}
  {\path{doi:https://doi.org/10.1142/p614}}.

\bibitem{Fabrizio2014206}
M.~Fabrizio, Fractional rheological models for thermomechanical systems.
  dissipation and free energies, Fractional Calculus and Applied Analysis
  17~(1) (2014) 206--223.
\newblock \href {http://dx.doi.org/https://doi.org/10.2478/s13540-014-0163-7}
  {\path{doi:https://doi.org/10.2478/s13540-014-0163-7}}.

\bibitem{Tarasov20161650128}
V.~E. Tarasov, Poiseuille equation for steady flow of fractal fluid,
  International Journal of Modern Physics B 30~(22) (2016) 1650128.
\newblock \href {http://dx.doi.org/https://doi.org/10.1142/S0217979216501289}
  {\path{doi:https://doi.org/10.1142/S0217979216501289}}.

\bibitem{Bouchendouka2022582}
A.~Bouchendouka, Z.~E.~A. Fellah, Z.~Larbi, N.~O. Ongwen, E.~Ogam, M.~Fellah,
  C.~Depollier, Flow of a self-similar non-{Newtonian} fluid using fractal
  dimensions, Fractal and Fractional 6~(10) (2022) 582.
\newblock \href {http://dx.doi.org/https://doi.org/10.3390/fractalfract6100582}
  {\path{doi:https://doi.org/10.3390/fractalfract6100582}}.

\bibitem{Oldham2006}
K.~B. Oldham, J.~Spanier, The Fractional Calculus: Theory and Applications of
  Differentiation and Integration to Arbitrary Order, Dover Books on
  Mathematics, Dover Publications, 2006.

\bibitem{Miller1993}
K.~S. Miller, B.~Ross, An Introduction to the Fractional Calculus and
  Fractional Differential Equations, Wiley, New York, 1993.

\bibitem{Samko1993}
S.~G. Samko, A.~A. Kilbas, O.~I. Marichev, Fractional Integrals and
  Derivatives: Theory and Applications, 1st Edition, Gordon and Breach Science
  Publishers, 1993.

\bibitem{Podlubny1998}
I.~Podlubny, V.~T.~E. Kenneth, Fractional Differential Equations: An
  Introduction to Fractional Derivatives, Fractional Differential Equations, to
  Methods of Their Solution and Some of Their Applications, 1st Edition,
  Mathematics in Science and Engineering 198, Academic Press, 1998.

\bibitem{Uchaikin2008500}
V.~V. Uchaikin, Fractional Derivatives Method, Artishock-Press, Uljanovsk,
  2008.

\bibitem{Tarasov2010}
V.~E. Tarasov, Fractional Dynamics: Applications of Fractional Calculus to
  Dynamics of Particles, Fields and Media, 1st Edition, Nonlinear Physical
  Science, Springer Berlin Heidelberg, 2010.

\bibitem{Caputo201573}
M.~Caputo, M.~Fabrizio, A new definition of fractional derivative without
  singular kernel, Progress in Fractional Differentiation \& Applications 1~(2)
  (2015) 73--85.
\newblock \href {http://dx.doi.org/https://doi.org/10.12785/pfda/010201}
  {\path{doi:https://doi.org/10.12785/pfda/010201}}.

\bibitem{Sahimi1998213}
M.~Sahimi, Non-linear and non-local transport processes in heterogeneous media:
  from long-range correlated percolation to fracture and materials breakdown,
  Physics Reports 306~(4--6) (1998) 213--395.
\newblock \href
  {http://dx.doi.org/https://doi.org/10.1016/S0370-1573(98)00024-6}
  {\path{doi:https://doi.org/10.1016/S0370-1573(98)00024-6}}.

\bibitem{Korosak20071}
D.~Koro\u{s}ak, B.~Cvikl, J.~Kramer, R.~Jecl, A.~Prapotnik, Fractional calculus
  applied to the analysis of spectral electrical conductivity of clay–water
  system, Journal of Contaminant Hydrology 92~(1–-2) (2007) 1--9.
\newblock \href
  {http://dx.doi.org/http://dx.doi.org/10.1016/j.jconhyd.2006.11.005}
  {\path{doi:http://dx.doi.org/10.1016/j.jconhyd.2006.11.005}}.

\bibitem{Hobbie2007}
R.~K. Hobbie, B.~J. Roth, {Intermediate Physics for Medicine and Biology}, 4th
  Edition, Springer-Verlag, New York, 2007.

\bibitem{Compte19977277}
A.~Compte, R.~Metzler, {The generalized Cattaneo equation for the description
  of anomalous transport processes}, Journal of Physics A: Mathematical and
  General 30~(21) (1997) 7277--7289.
\newblock \href {http://dx.doi.org/https://doi.org/10.1088/0305-4470/30/21/006}
  {\path{doi:https://doi.org/10.1088/0305-4470/30/21/006}}.

\bibitem{Metzler20001}
R.~Metzler, J.~Klafter, The random walk's guide to anomalous diffusion: a
  fractional dynamics approach, Physics Reports 339~(1) (2000) 1--77.
\newblock \href
  {http://dx.doi.org/http://dx.doi.org/10.1016/S0370-1573(00)00070-3}
  {\path{doi:http://dx.doi.org/10.1016/S0370-1573(00)00070-3}}.

\bibitem{Hilfer19951475}
R.~Hilfer, Fractional dynamics, irreversibility and ergodicity breaking, Chaos,
  Solitons \& Fractals 5~(8) (1995) 1475--1484.
\newblock \href
  {http://dx.doi.org/http://dx.doi.org/10.1016/0960-0779(95)00027-2}
  {\path{doi:http://dx.doi.org/10.1016/0960-0779(95)00027-2}}.

\bibitem{Hilfer20003914}
R.~Hilfer, {Fractional Diffusion Based on {Riemann-Liouville} Fractional
  Derivatives}, The Journal of Physical Chemistry B 104~(16) (2000) 3914--3917.
\newblock \href {http://dx.doi.org/https://doi.org/10.1021/jp9936289}
  {\path{doi:https://doi.org/10.1021/jp9936289}}.

\bibitem{Hilfer2000}
R.~Hilfer, Fractional Time Evolution, World Scientific, Singapore, New Jersey,
  London, Hong Kong, 2000, Ch.~II, pp. 87--130.

\bibitem{Kosztolowicz2005041105}
T.~Koszto\l{}owicz, K.~Dworecki, S.~Mr\'owczy\ifmmode~\acute{n}\else
  \'{n}\fi{}ski, Measuring subdiffusion parameters, Phys. Rev. E 71~(4) (2005)
  041105.
\newblock \href {http://dx.doi.org/https://doi.org/10.1103/PhysRevE.71.041105}
  {\path{doi:https://doi.org/10.1103/PhysRevE.71.041105}}.

\bibitem{Kosztolowicz2015P10021}
T.~Koszto\l{}owicz, Subdiffusive random walk in a membrane system: the
  generalized method of images approach, Journal of Statistical Mechanics:
  Theory and Experiment 2015~(10) (2015) P10021.

\bibitem{Bisquert20002287}
J.~Bisquert, G.~Garcia-Belmonte, F.~Fabregat-Santiago, N.~S. Ferriols,
  P.~Bogdanoff, E.~C. Pereira, {Doubling Exponent Models for the Analysis of
  Porous Film Electrodes by Impedance. Relaxation of TiO$_2$ Nanoporous in
  Aqueous Solution}, The Journal of Physical Chemistry B 104~(10) (2000)
  2287--2298.
\newblock \href {http://dx.doi.org/https://doi.org/10.1021/jp993148h}
  {\path{doi:https://doi.org/10.1021/jp993148h}}.

\bibitem{Bisquert2001112}
J.~Bisquert, A.~Compte, Theory of the electrochemical impedance of anomalous
  diffusion, Journal of Electroanalytical Chemistry 499~(1) (2001) 112--120.
\newblock \href
  {http://dx.doi.org/http://dx.doi.org/10.1016/S0022-0728(00)00497-6}
  {\path{doi:http://dx.doi.org/10.1016/S0022-0728(00)00497-6}}.

\bibitem{Kosztolowicz2009055004}
T.~Koszto\l{}owicz, K.~D. Lewandowska, Hyperbolic subdiffusive impedance,
  Journal of Physics A: Mathematical and Theoretical 42~(5) (2009) 055004.
\newblock \href
  {http://dx.doi.org/https://doi.org/10.1088/1751-8113/42/5/055004}
  {\path{doi:https://doi.org/10.1088/1751-8113/42/5/055004}}.

\bibitem{Pyanylo201484}
Y.~D. Pyanylo, M.~G. Prytula, N.~M. Prytula, N.~B. Lopuh, Models of mass
  transfer in gas transmission systems, Mathematical Modeling and Computing
  1~(1) (2014) 84--96.
\newblock \href {http://dx.doi.org/https://doi.org/10.23939/mmc2014.01.084}
  {\path{doi:https://doi.org/10.23939/mmc2014.01.084}}.

\bibitem{Kostrobij2016093301}
P.~Kostrobij, B.~Markovych, O.~Viznovych, M.~Tokarchuk, {Generalized diffusion
  equation with fractional derivatives within Renyi statistics}, Journal of
  Mathematical Physics 57~(9) (2016) 093301.
\newblock \href {http://dx.doi.org/https://doi.org/10.1063/1.4962159}
  {\path{doi:https://doi.org/10.1063/1.4962159}}.

\bibitem{Kostrobij2016163}
P.~Kostrobij, B.~Markovych, O.~Viznovych, M.~Tokarchuk, Generalized
  electrodiffusion equation with fractality of space-time, Mathematical
  Modeling and Computing 3~(2) (2016) 163--172.
\newblock \href {http://dx.doi.org/https://doi.org/10.23939/mmc2016.02.163}
  {\path{doi:https://doi.org/10.23939/mmc2016.02.163}}.

\bibitem{Glushak201857}
P.~A. Glushak, B.~B. Markiv, M.~V. Tokarchuk,
  \href{https://doi.org/10.1134/S0040577918010051}{Zubarev’s nonequilibrium
  statistical operator method in the generalized statistics of multiparticle
  systems}, Theoretical and Mathematical Physics 194~(1) (2018) 57--73.
\newblock \href {http://dx.doi.org/https://doi.org/10.1134/S0040577918010051}
  {\path{doi:https://doi.org/10.1134/S0040577918010051}}.
\newline\urlprefix\url{https://doi.org/10.1134/S0040577918010051}

\bibitem{Kostrobij201963}
P.~P. Kostrobij, B.~M. Markovych, O.~V. Viznovych, M.~V. Tokarchuk,
  {Generalized transport equation with nonlocality of space–time. Zubarev’s NSO
  method}, Physica A: Statistical Mechanics and its Applications 514 (2019)
  63--70.
\newblock \href {http://dx.doi.org/https://doi.org/10.1016/j.physa.2018.09.051}
  {\path{doi:https://doi.org/10.1016/j.physa.2018.09.051}}.

\bibitem{Kostrobij201958}
P.~Kostrobij, B.~Markovych, O.~Viznovych, I.~Zelinska, M.~Tokarchuk,
  \href{https://doi.org/10.23939/mmc2019.01.058}{Generalized cattaneo–maxwell
  diffusion equation with fractional derivatives. dispersion relations},
  Mathematical Modeling and Computing 6~(1) (2019) 58--68.
\newblock \href {http://dx.doi.org/https://doi.org/10.23939/mmc2019.01.058}
  {\path{doi:https://doi.org/10.23939/mmc2019.01.058}}.
\newline\urlprefix\url{https://doi.org/10.23939/mmc2019.01.058}

\bibitem{Grygorchak2015e}
I.~I. Grygorchak, P.~P. Kostrobij, I.~V. Stasjuk, M.~V. Tokarchuk, O.~V.
  Velychko, F.~O. Ivaschyshyn, B.~M. Markovych, Fizichni procesy ta ih
  mikroskopichni modeli v periodychnyh neorganichno/organichnih klatratah,
  Rastr-7, Lviv, 2015.

\bibitem{Kostrobij2015154}
P.~P. Kostrobij, I.~I. Grygorchak, F.~O. Ivaschyshyn, B.~M. Markovych, O.~V.
  Viznovych, M.~V. Tokarchuk, Mathematical modeling of subdiffusion impedance
  in multilayer nanostructures, Mathematical Modeling and Computing 2~(2)
  (2015) 154--159.
\newblock \href {http://dx.doi.org/https://doi.org/10.23939/mmc2015.02.154}
  {\path{doi:https://doi.org/10.23939/mmc2015.02.154}}.

\bibitem{Grygorchak2017185501}
I.~I. Grygorchak, F.~O. Ivashchyshyn, M.~V. Tokarchuk, N.~T. Pokladok, O.~V.
  Viznovych, {Modification of properties of
  GaSe$\langle\beta$-cyclodexterin$\langle$FeSO$_4\rangle\rangle$ Clathrat by
  synthesis in superposed electric and light-wave fields}, Journal of Applied
  Physics 121~(18) (2017) 185501.
\newblock \href {http://dx.doi.org/https://doi.org/10.1063/1.4983097}
  {\path{doi:https://doi.org/10.1063/1.4983097}}.

\bibitem{Kostrobij20184099}
P.~Kostrobij, I.~Grygorchak, F.~Ivashchyshyn, B.~Markovych, O.~Viznovych,
  M.~Tokarchuk, Generalized electrodiffusion equation with fractality of
  space–time: Experiment and theory, The Journal of Physical Chemistry A
  122~(16) (2018) 4099--4110.
\newblock \href {http://dx.doi.org/https://doi.org/10.1021/acs.jpca.8b00188}
  {\path{doi:https://doi.org/10.1021/acs.jpca.8b00188}}.

\bibitem{Sahimi2011}
M.~Sahimi, {Flow and Transport in Porous Media and Fractured Rock: From
  Classical Methods to Modern Approaches}, John Wiley \& Sons, Ltd, 2011.
\newblock \href {http://dx.doi.org/https://doi.org/10.1002/9783527636693}
  {\path{doi:https://doi.org/10.1002/9783527636693}}.

\bibitem{Dietrich2005}
P.~Dietrich, R.~Helmig, M.~Sauter, H.~H\"{o}tzl, J.~K\"{o}ngeter, G.~Teutsch
  (Eds.), {Flow and Transport in Fractured Porous Media}, Springer Verlag,
  2005.

\bibitem{Winters1994}
M.~E. Winters, {Basic Clinical Pharmacokinetics}, Applied Therapeutics, Inc.,
  Vancouver, British Columbia, 1994.

\bibitem{Almeida19995486}
M.~P. Almeida, J.~S. Andrade, S.~V. Buldyrev, F.~S.~A. Cavalcante, H.~E.
  Stanley, B.~Suki, Fluid flow through ramified structures, Phys. Rev. E 60~(5)
  (1999) 5486--5494.
\newblock \href {http://dx.doi.org/https://doi.org/10.1103/PhysRevE.60.5486}
  {\path{doi:https://doi.org/10.1103/PhysRevE.60.5486}}.

\bibitem{Knudsen2002056310}
H.~A. Knudsen, A.~Hansen, Relation between pressure and fractional flow in
  two-phase flow in porous media, Phys. Rev. E 65~(5) (2002) 056310.
\newblock \href {http://dx.doi.org/https://doi.org/10.1103/PhysRevE.65.056310}
  {\path{doi:https://doi.org/10.1103/PhysRevE.65.056310}}.

\bibitem{Huinink2002046301}
H.~P. Huinink, M.~A.~J. Michels, Influence of buoyancy on drainage of a fractal
  porous medium, Phys. Rev. E 66~(4) (2002) 046301.
\newblock \href {http://dx.doi.org/https://doi.org/10.1103/PhysRevE.66.046301}
  {\path{doi:https://doi.org/10.1103/PhysRevE.66.046301}}.

\bibitem{Balankin2011036310}
A.~S. Balankin, S.~M. Gutierres, D.~S. Ochoa, J.~P.~n. Ortiz, B.~E.
  Elizarraraz, C.~L. Mart\'{\i}nez-Gonz\'alez, Slow kinetics of water escape
  from randomly folded foils, Phys. Rev. E 83~(3) (2011) 036310.
\newblock \href {http://dx.doi.org/https://doi.org/10.1103/PhysRevE.83.036310}
  {\path{doi:https://doi.org/10.1103/PhysRevE.83.036310}}.

\bibitem{Lopez2003056314}
E.~L\'opez, S.~V. Buldyrev, N.~V. Dokholyan, L.~Goldmakher, S.~Havlin, P.~R.
  King, H.~E. Stanley, Postbreakthrough behavior in flow through porous media,
  Phys. Rev. E 67~(5) (2003) 056314.
\newblock \href {http://dx.doi.org/https://doi.org/10.1103/PhysRevE.67.056314}
  {\path{doi:https://doi.org/10.1103/PhysRevE.67.056314}}.

\bibitem{Stanley200117}
H.~E. Stanley, J.~S. Andrade, Physics of the cigarette filter: fluid flow
  through structures with randomly-placed obstacles, Physica A: Statistical
  Mechanics and its Applications 295~(1) (2001) 17--30.
\newblock \href
  {http://dx.doi.org/https://doi.org/10.1016/S0378-4371(01)00140-6}
  {\path{doi:https://doi.org/10.1016/S0378-4371(01)00140-6}}.

\bibitem{Chen2009026301}
Y.~Chen, C.~Zhang, M.~Shi, G.~P. Peterson, Role of surface roughness
  characterized by fractal geometry on laminar flow in microchannels, Phys.
  Rev. E 80~(2) (2009) 026301.
\newblock \href {http://dx.doi.org/https://doi.org/10.1103/PhysRevE.80.026301}
  {\path{doi:https://doi.org/10.1103/PhysRevE.80.026301}}.

\bibitem{Tian2006287}
J.~Tian, D.-k. Tong, The flow analysis of fiuids in fractal reservoir with the
  fractional derivative, Journal of Hydrodynamics 18 (2006) 287--293.
\newblock \href
  {http://dx.doi.org/https://doi.org/10.1016/S1001-6058(06)60005-X}
  {\path{doi:https://doi.org/10.1016/S1001-6058(06)60005-X}}.

\bibitem{Wheatcraft20081377}
S.~W. Wheatcraft, M.~M. Meerschaert, Fractional conservation of mass, Advances
  in Water Resources 31~(10) (2008) 1377--1381.

\bibitem{Tarasov2005167}
V.~E. Tarasov, Continuous medium model for fractal media, Physics Letters A
  336~(2) (2005) 167--174.
\newblock \href
  {http://dx.doi.org/https://doi.org/10.1016/j.physleta.2005.01.024}
  {\path{doi:https://doi.org/10.1016/j.physleta.2005.01.024}}.

\bibitem{Tarasov2005286}
V.~E. Tarasov, Fractional hydrodynamic equations for fractal media, Annals of
  Physics 318~(2) (2005) 286--307.
\newblock \href {http://dx.doi.org/http://dx.doi.org/10.1016/j.aop.2005.01.004}
  {\path{doi:http://dx.doi.org/10.1016/j.aop.2005.01.004}}.

\bibitem{Ostoja20071085}
M.~Ostoja-Starzewski, Towards thermomechanics of fractal media, Zeitschrift
  f\"{u}r angewandte Mathematik und Physik 58 (2007) 1085--1096.
\newblock \href {http://dx.doi.org/https://doi.org/10.1007/s00033-007-7027-5}
  {\path{doi:https://doi.org/10.1007/s00033-007-7027-5}}.

\bibitem{Li20092521}
J.~Li, M.~Ostoja-Starzewski, Fractal solids, product measures and fractional
  wave equations, Proceedings of the Royal Society A: Mathematical, Physical
  and Engineering Sciences 465~(2108) (2009) 2521--2536.
\newblock \href {http://dx.doi.org/10.1098/rspa.2009.0101}
  {\path{doi:10.1098/rspa.2009.0101}}.

\bibitem{Balankin2012025302}
A.~S. Balankin, B.~E. Elizarraraz, Hydrodynamics of fractal continuum flow,
  Phys. Rev. E 85~(2) (2012) 025302.
\newblock \href {http://dx.doi.org/10.1103/PhysRevE.85.025302}
  {\path{doi:10.1103/PhysRevE.85.025302}}.

\bibitem{Balankin2012056314}
A.~S. Balankin, B.~E. Elizarraraz, Map of fluid flow in fractal porous medium
  into fractal continuum flow, Phys. Rev. E 85~(5) (2012) 056314.
\newblock \href {http://dx.doi.org/10.1103/PhysRevE.85.056314}
  {\path{doi:10.1103/PhysRevE.85.056314}}.

\bibitem{Li2013057001}
J.~Li, M.~Ostoja-Starzewski, Comment on ``hydrodynamics of fractal continuum
  flow'' and ``map of fluid flow in fractal porous medium into fractal
  continuum flow'', Phys. Rev. E 88~(5) (2013) 057001.
\newblock \href {http://dx.doi.org/10.1103/PhysRevE.88.057001}
  {\path{doi:10.1103/PhysRevE.88.057001}}.

\bibitem{Li202020190288}
J.~Li, M.~Ostoja-Starzewski, Thermo-poromechanics of fractal media,
  Philosophical Transactions of the Royal Society A: Mathematical, Physical and
  Engineering Sciences 378~(2172) (2020) 20190288.
\newblock \href {http://dx.doi.org/10.1098/rsta.2019.0288}
  {\path{doi:10.1098/rsta.2019.0288}}.

\bibitem{Zubarev19811509}
D.~N. Zubarev, Modern methods of the statistical theory of nonequilibrium
  processes, Journal of Soviet Mathematics 16~(6) (1981) 1509--1571.
\newblock \href {http://dx.doi.org/https://doi.org/10.1007/BF01091712}
  {\path{doi:https://doi.org/10.1007/BF01091712}}.

\bibitem{Zubarev20021}
D.~N. Zubarev, V.~G. Morozov, G.~R\"{o}pke, Statistical mechanics of
  nonequilibrium processes, Vol.~1, Fizmatlit, 2002.

\bibitem{Zubarev20022}
D.~N. Zubarev, V.~G. Morozov, G.~R\"{o}pke, Statistical mechanics of
  nonequilibrium processes, Vol.~2, Fizmatlit, 2002.

\bibitem{Tarasov2004123}
V.~E. Tarasov, {Fractional generalization of Liouville equations}, Chaos 14~(1)
  (2004) 123--127.
\newblock \href {http://dx.doi.org/http://dx.doi.org/10.1063/1.1633491}
  {\path{doi:http://dx.doi.org/10.1063/1.1633491}}.

\bibitem{Tarasov200517}
V.~E. Tarasov, {Fractional Liouville and BBGKI equations}, Journal of Physics:
  Conference Series 7~(1) (2005) 17--33.
\newblock \href {http://dx.doi.org/https://doi.org/10.1088/1742-6596/7/1/002}
  {\path{doi:https://doi.org/10.1088/1742-6596/7/1/002}}.

\bibitem{Tarasov2005011102}
V.~E. Tarasov, Fractional systems and fractional {Bogoliubov} hierarchy
  equations, Phys. Rev. E 71~(1) (2005) 011102.
\newblock \href {http://dx.doi.org/https://doi.org/10.1103/PhysRevE.71.011102}
  {\path{doi:https://doi.org/10.1103/PhysRevE.71.011102}}.

\bibitem{Tarasov2006033108}
V.~E. Tarasov, Fractional statistical mechanics, Chaos 16~(3) (2006) 033108.
\newblock \href {http://dx.doi.org/http://dx.doi.org/10.1063/1.2219701}
  {\path{doi:http://dx.doi.org/10.1063/1.2219701}}.

\bibitem{Tarasov2006341}
V.~E. Tarasov, {Transport equations from Liouville equations for fractional
  systems}, International Journal of Modern Physics B 20~(3) (2006) 341--353.
\newblock \href {http://dx.doi.org/https://doi.org/10.1142/S0217979206033267}
  {\path{doi:https://doi.org/10.1142/S0217979206033267}}.

\bibitem{El-Nabulsi2019449}
R.~A. El-Nabulsi, Fractional {Navier–Stokes} equation from fractional velocity
  arguments and its implications in fluid flows and microfilaments,
  International Journal of Nonlinear Sciences and Numerical Simulation 20~(3-4)
  (2019) 449--459.
\newblock \href {http://dx.doi.org/https://doi.org/10.1515/ijnsns-2018-0253}
  {\path{doi:https://doi.org/10.1515/ijnsns-2018-0253}}.

\bibitem{Kathleen20012203}
K.~Cottrill-Shepherd, M.~Naber, Fractional differential forms, Journal of
  Mathematical Physics 42~(5) (2001) 2203--2212.
\newblock \href {http://dx.doi.org/http://dx.doi.org/10.1063/1.1364688}
  {\path{doi:http://dx.doi.org/10.1063/1.1364688}}.

\bibitem{Tarasov20082756}
V.~E. Tarasov,
  \href{http://www.sciencedirect.com/science/article/pii/S0003491608000596}{Fractional
  vector calculus and fractional maxwell’s equations}, Annals of Physics
  323~(11) (2008) 2756--2778.
\newblock \href {http://dx.doi.org/https://doi.org/10.1016/j.aop.2008.04.005}
  {\path{doi:https://doi.org/10.1016/j.aop.2008.04.005}}.
\newline\urlprefix\url{http://www.sciencedirect.com/science/article/pii/S0003491608000596}

\bibitem{Tarasov2011e}
V.~E. Tarasov, Modeli teoreticheskoj fiziki s integro-differencirovaniem
  drobnogo porjadka, Izhevskij institut komp'juternyh issledovanij, Izhevskij
  institut komp'juternyh issledovanij, Moskva--Izhevsk, 2011.

\bibitem{Mainardi1997}
F.~Mainardi, Fractional Calculus, Springer, Vienna, 1997.
\newblock \href {http://dx.doi.org/https://doi.org/10.1007/978-3-7091-2664-6_7}
  {\path{doi:https://doi.org/10.1007/978-3-7091-2664-6_7}}.

\bibitem{Caputo1971134}
M.~Caputo, F.~Mainardi, A new dissipation model based on memory mechanism, Pure
  and Applied Geophysics 91~(1) (1971) 134--147.
\newblock \href {http://dx.doi.org/https://doi.org/10.1007/BF00879562}
  {\path{doi:https://doi.org/10.1007/BF00879562}}.

\bibitem{Akcasu1970962}
A.~Z. Akcasu, E.~Daniels, Fluctuation analysis in simple fluids, Phys. Rev. A
  2~(3) (1970) 962--975.
\newblock \href {http://dx.doi.org/https://doi.org/10.1103/PhysRevA.2.962}
  {\path{doi:https://doi.org/10.1103/PhysRevA.2.962}}.

\bibitem{Schepper1988271}
I.~M. de~Schepper, E.~G.~D. Cohen, C.~Bruin, J.~C. van Rijs, W.~Montfrooij,
  L.~A. de~Graaf, Hydrodynamic time correlation functions for a {Lennard-Jones}
  fluid, Phys. Rev. A 38~(1) (1988) 271--287.
\newblock \href {http://dx.doi.org/https://doi.org/10.1103/PhysRevA.38.271}
  {\path{doi:https://doi.org/10.1103/PhysRevA.38.271}}.

\bibitem{Mryglod1995235}
I.~M. Mryglod, I.~P. Omelyan, M.~V. Tokarchuk, {Generalized collective modes
  for the Lennard-Jones fluid}, Molecular Physics 84~(2) (1995) 235--259.
\newblock \href {http://dx.doi.org/https://doi.org/10.1080/00268979500100181}
  {\path{doi:https://doi.org/10.1080/00268979500100181}}.

\bibitem{Ignatyuk1999295}
V.~V. Ignatyuk, I.~M. Mryglod, M.~V. Tokarchuk, {On the theory of dynamic
  properties of semiquantum helium}, Low Temperature Physics 25~(5) (1999)
  295--302.
\newblock \href {http://dx.doi.org/https://doi.org/10.1063/1.593742}
  {\path{doi:https://doi.org/10.1063/1.593742}}.

\bibitem{Ignatyuk1999857}
V.~V. Ignatyuk, M.~V. Tokarchuk, I.~M. Mryglod, Time correlation functions and
  generalized transport coefficients of semiquantum helium, Low Temperature
  Physics 25~(11) (1999) 857--863.
\newblock \href {http://dx.doi.org/https://doi.org/10.1063/1.593830}
  {\path{doi:https://doi.org/10.1063/1.593830}}.

\bibitem{Zubarev1987164}
D.~N. Zubarev, M.~V. Tokarchuk, Nonequilibrium statistical hydrodynamics of
  ionic systems, Theoretical and Mathematical Physics volume 70 (1987)
  164--178.
\newblock \href {http://dx.doi.org/https://doi.org/10.1007/BF01039207}
  {\path{doi:https://doi.org/10.1007/BF01039207}}.

\bibitem{Tokarchuk2022440}
M.~V. Tokarchuk, To the kinetic theory of dense gases and liquids. calculation
  of quasi-equilibrium particle distribution functions by the method of
  collective variables, Mathematical Modeling and Computing 9~(2) (2022)
  440--458.
\newblock \href {http://dx.doi.org/https://doi.org/10.23939/mmc2022.02.440}
  {\path{doi:https://doi.org/10.23939/mmc2022.02.440}}.

\end{thebibliography}

\end{document}